\RequirePackage{color}
\documentclass{PoS}

\usepackage{graphicx,cite}
\usepackage{color}
\usepackage{colortbl}
\usepackage[]{xcolor}
\usepackage{afterpage}
\input macros.sty

\title{Low-energy particle physics and chiral extrapolations}

\ShortTitle{Chiral extrapolations}

\author{\speaker{Hartmut Wittig}
        \\
        Institut f\"ur Kernphysik and Helmholtz Institute Mainz,
        University of Mainz, Becher Weg 45, 55099 Mainz, Germany\\
        E-mail: \email{wittig@kph.uni-mainz.de}}

\abstract{
\vspace{-10.5cm}
  In this review I discuss the r\^ole of chiral extrapolations
  for the determination of several phenomenologically relevant
  quantities, including light quark masses, meson decay constants and
  the axial charge of the nucleon. In particular, I investigate
  whether chiral extrapolations are sufficiently controlled in order
  to rightfully claim the accuracy which is quoted in recent
  compilations of these quantities. While this is the case for the
  masses of the light quarks and the ratio $\fK/f_\pi$ of decay
  constants, small inconsistencies in the chiral and continuum
  behaviour of individual decay constants $\fK$ and $f_\pi$, as well
  as the hadronic radii $r_0, r_1$ remain and must be clarified. In
  the case of the nucleon axial charge, $\gA$, the chiral behaviour is
  still poorly understood due to the presence of other systematic
  effects.
\vspace{-15.5cm}{\begin{flushright} {\tt MKPH-T-11-25\\ HIM-2011-12
              }\end{flushright}}
}

\FullConference{ The XXIX International Symposium on Lattice Field
  Theory - Lattice 2011\\ 
  July 10-16, 2011\\
  Squaw Valley, Lake Tahoe, California}

\begin{document}

\section{Introduction \label{s1intro}}

Lattice calculations are becoming increasingly important for particle
physics phenomenology. They address and quantify the ``hadronic
uncertainties'' which still afflict many quantities that constrain the
validity of the Standard
Model\,\cite{FLAG2010,LLVdW2009,Lunghi:lat11,Davies:lat11,Mawhinney:lat11,Renner:lat11}.
Almost every talk on lattice QCD delivered to a more general audience
during the past 10--15 years contained the phrase that ``lattice
calculations are performed at unphysical quark masses.'' What we
usually mean by this statement is that for any given discretisation,
there is {\it a priori} no way of knowing which values of the bare
quark masses correspond to those of the physical quarks. In most, if
not all, cases it turns out that the physical light quark masses lie
outside the regime which is directly accessible using the currently
available algorithms and machines. Likewise, the physical values of
the heavy quarks are dangerously close to, if not above, the
affordable cutoff scale.

Chiral extrapolations are thus required in order to make contact with
the physical light quark masses. Chiral Perturbation Theory (ChPT)
provides theoretical constraints on the quark mass dependence of
observables, based on the underlying dynamics associated with chiral
symmetry breaking. The following quotation from the FLAG
report\,\cite{FLAG2010} serves as a reminder that improving the
control over the chiral behaviour is mandatory in order to make
further progress:
\begin{center}
\parbox[t]{14cm}{
{\sl ``Although [light quark masses] are decreasing very significantly
  with time [\ldots] it remains true that [the chiral] extrapolation is
  one of the most significant sources of systematic error.''}
}
\end{center}
Some collaborations have produced lattice data at or even below the
physical pion mass\,\cite{BMW2010-1,PACS-CS2009} which may render
extrapolations guided by ChPT soon obsolete. Instead, one can resort
to some sort of analytic {\it ansatz} to interpolate lattice data to
the physical point. However, even if ChPT were not required to perform
chiral extrapolations, a comparison of the quark mass dependence
determined on the lattice with the predictions of an effective theory
would provide useful information, since it allows for the
determination of the low-energy constants (LECs) which parameterise
ChPT. Furthermore, as simulation algorithms still show a significant
increase in computational cost when the light quark masses are tuned
to their physical values, it is still difficult to disentangle the
chiral behaviour from systematic effects arising from finite volume
and/or coarse lattice spacings.

The vast majority of lattice estimates for phenomenologically relevant
quantities is still dominated by systematic errors. In this review I
try to investigate whether chiral extrapolations are sufficiently well
controlled in order to rightfully claim the accuracy which is quoted
in the recent compilations. Here I shall focus on three different
types of observables: Lattice estimates of the light quark masses are
discussed in the next section. In section~\ref{sec:s3decay} I will
study the systematics of chiral fits applied to meson decay
constants. Section~\ref{sec:s4axial} contains a discussion of the
chiral behaviour of the axial charge of the nucleon. Summary and
conclusions are provided in section~\ref{sec:s5summ}.
For the purpose of this review all lattice results are taken at face
value. Issues such as ``rooting'', induced non-localities, or the
freezing of topology will not be discussed here.

\section{Light quark masses} \label{sec:s2quark}

Quark masses are fundamental parameters of the Standard Model whose
values determine many important quantities in particle
phenomenology. The recent compilation of lattice results for the light
quark masses and their conversion into ``global'' averages in the FLAG
report\,\cite{FLAG2010} is based on a set of ``quality
criteria''. Using a simple colour code, they are meant to assess the
quality of a given calculation regarding a number of different
systematic effects. Obviously, these criteria must be adjusted over
time, in order to reflect the true state-of-the-art. In the current
FLAG review, a green star (\tbg) is awarded if the systematic error is
``convincingly shown to be under control''. An amber ball ({\tbo})
signifies that a ``reasonable attempt'' at estimating a particular
systematic error has been made. Finally, a red box ({\tbr}) indicates
that no attempt was undertaken to quantify a systematic effect. To set
the scene for the discussions to follow, the criteria for the colour
code referring to chiral extrapolations and the related finite-volume
effects are repeated here:
\begin{center}
\parbox[t]{15cm}{
\parbox[b]{7cm}{
  Chiral extrapolation:
  \begin{itemize}\itemsep 0pt
  \item[{\small\tbg}] $m_\pi^{\sf min} < 250\,\MeV$
  \item[\tbo] $250\,\MeV\leq m_\pi^{\sf min} \leq 400\,\MeV$
  \item[{\small\tbr}] $m_\pi^{\sf min} > 400\,\MeV$
  \end{itemize}
}
\hfill
\parbox[b]{7cm}{
  Finite-volume effects: 
  \begin{itemize}\itemsep 0pt
  \item[{\small\tbg}] $m_\pi^{\sf min}L > 4$ or at least 3 volumes
  \item[\tbo] $m_\pi^{\sf min}L > 3$ and at least 2 volumes
  \item[{\small\tbr}] otherwise, or if $(L_{\sf min} < 2\,\fm)$
  \end{itemize}
}
}
\vspace{-0.4cm}
\end{center}
\begin{figure}
\begin{center}
\includegraphics[height=7.5cm]{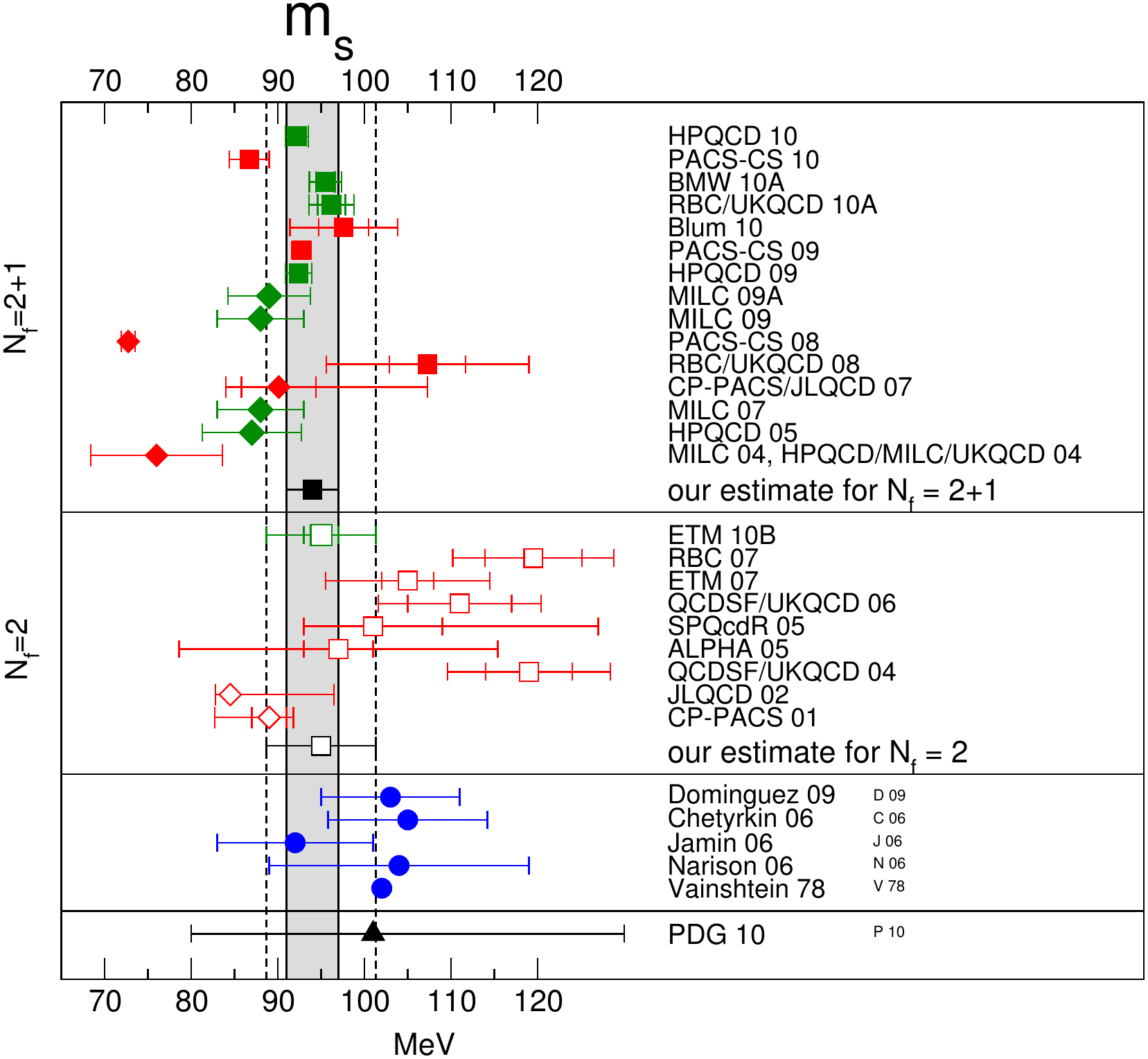}
\caption{\label{fig:FLAGms} Results for the strange quark mass in the
  $\msbar$-scheme at $2\,\GeV$, obtained in lattice QCD with $\Nf=2+1$ and
  $\Nf=2$ flavours of dynamical quarks\,\cite{FLAG2010}. Green points
  represent lattice results which are free of any red tags according to the
  FLAG criteria. Blue circles denote the results from sum rule
  calculations. The grey band and the vertical dotted lines denote the global
  estimate for the $\Nf=2+1$ and two-flavour theory, respectively. The PDG
  estimate is shown at the bottom.}
\vspace{-0.2cm}
\end{center}
\end{figure}

Moreover, the FLAG rules stipulate that results which are classified
with at least one red tag and/or without a journal reference be
excluded from global estimates. As an example we show the compilation
of results for the strange quark mass from the FLAG report in
Fig.\,\ref{fig:FLAGms}. One observes that lattice estimates for $m_s$
obtained with $\Nf=2+1$ flavours of dynamical quarks appear to be
somewhat smaller compared to the two-flavour theory, although this may
be attributed to the fact that the results for $\Nf=2$ are typically
older and may be more strongly affected by other systematic effects. A
striking feature of the plot is that lattice estimates are broadly
consistent, despite the fact that they have been obtained for several
different discretisations of the quark action. Moreover, the quoted
uncertainties are much smaller than those attributed to sum rule
results and the PDG average {PDG2010}. The FLAG report quotes the
following global estimates, based on the results of
refs.\,\cite{MILC09A,RBC-UKQCD10A} and\,\cite{HPQCD10}:
\be
  m_{ud}^{\msbar}(2\,\GeV)=3.43(11)\,\MeV,\quad
  m_{s}^{\msbar}(2\,\GeV)=94(3)\,\MeV,\quad m_s/m_{ud}=27.4\pm0.4.
\label{eq:quark}
\ee
This high level of accuracy raises the question whether systematic
effects, in particularly those associated with the chiral
extrapolation, are indeed controlled.

\begin{table}
\begin{center}
\begin{tabular}{l c c c l c c}
\hline\hline
  & Ref.  & $a\,[\fm]$ & $m_\pi^{\sf{min}}\,[\MeV]$ & $m_\pi^{\sf{min}}L$ &
    $m_s^{\msbar}\,[\MeV]$ & $m_s/m_{ud}$ \\
\hline
  & \cite{Aoki:2008sm} & & & & $72.7\pm0.8\phantom{{}^\dagger}$
  & $28.8\pm0.4\phantom{{}^\dagger}$ \\
\rb{PACS-CS} & \cite{PACS-CS2009,Aoki:2010wm}
  & \rb{0.09} & \rb{156} & \rb{~2.3} & ${\red{86.7\pm2.3}}^\dagger$ &
$31.2\pm2.7^\dagger$ \\
\hline
BMW & \cite{BMW2010-2,BMW2010-1} & $\begin{array}{c} 0.116 \\ 0.093 \\ 0.077
  \\ 0.065 \\ 0.054  \end{array}$ & $\begin{array}{c} 136 \\ 131 \\120 \\182
  \\219 \end{array}$  & $\left.\begin{array}{c} 3.9 \\ 3.9 \\3.0 \\ 3.8
  \\3.8 \end{array}\quad \right\}$ &  ${\red{95.5\pm1.9}}$ & $27.53\pm0.22$ \\
\hline\hline
\end{tabular}
\caption{Results for the strange quark mass, $m_s^{\msbar}(2\,\GeV)$, and the
  ratio $m_s/m_{ud}$ from simulations near the physical pion mass. Results
  marked by a dagger were obtained using mass reweighting. Numbers in
  red are based on non-perturbatively determined renormalisation
  factors. \label{tab:quark}}
\end{center}
\vspace{-0.5cm}
\end{table}

We begin by discussing two recent calculations which do not rely on
chiral extrapolations. The PACS-CS
Collaboration\,\cite{Aoki:2008sm,PACS-CS2009,Aoki:2010wm} has used
non-perturbatively $\rmO(a)$ improved Wilson fermions and the Iwasaki
gauge action at a fixed value of the lattice spacing to determine the
light quark masses at the physical pion mass via mass reweighting. The
BMW Collaboration\,\cite{BMW2010-2,BMW2010-1} has performed
simulations with smeared tree-level improved Wilson quarks at pion
masses as low as 120\,\MeV. Quark masses were obtained by an
interpolation to the physical pion mass. Results and some simulation
details are shown in Table~\ref{tab:quark}.

At $m_\pi^{\sf min}=156\,\MeV$ the PACS-CS Collaboration is almost at
the physical point. Via a short chiral extrapolation, PACS-CS have
obtained the results shown in the first row of
Table\,\ref{tab:quark}\,\cite{Aoki:2008sm}. In a subsequent work they
have proceeded to simulate with hopping parameters
$(\kappa_{ud}^\ast,\kappa_s^\ast)_{\rm ext}$ which, according to the
chiral extrapolation of the results in\,\cite{Aoki:2008sm}, correspond
to the physical pion mass. In order to compensate the slight observed
mismatch between the targeted and actually measured pion mass, they
have reweighted their ensembles using the single-histogram
method\,\cite{Ferrenberg:1988yz}. The second line in
Table\,\ref{tab:quark} indicates that the resulting estimate for the
strange quark mass is quite different from the value obtained from an
extrapolation. However, this increase can be largely attributed to the
use of non-perturbative renormalisation factors in\,\cite{PACS-CS2009}
which were found to be 30\% larger than their perturbative
counterparts in ref.\,\cite{Aoki:2008sm}. This is consistent with the
observation that the difference between extrapolation and reweighting
is much less pronounced for the ratio $m_s/m_{ud}$ in which the
renormalisation factors cancel. One concludes that reweighting allows
one to avoid chiral extrapolations at the expense of incurring a
larger statistical error. Despite the large overall uncertainty, the
reweighted results by PACS-CS do not agree well with the global
estimates in\,\eq{eq:quark}. This may be explained by the presence of
other systematic errors, most notably lattice artefacts and
finite-volume effects, which are not yet sufficiently controlled.

For Wilson-type fermions, the BMW Collaboration has realised the
lowest pion masses so far, at least at the three coarsest lattice
spacings. However, their results for the light quark masses were not
included in the global estimates according to the FLAG rules,
since\,\cite{BMW2010-1,BMW2010-2} had not been published by the time
when ref.\,\cite{FLAG2010} was completed. BMW have performed a
comprehensive investigation of systematic effects, including studies
of the stability of their results under variations in the {\it ansatz}
for the fit function. To this end they compared chiral fits based on
either SU(2) ChPT at NLO or on a Taylor expansion, viz.
\be
   m_{ud}=\frac{m_\pi^2}{2{B}} \bigg\{
  1-\frac{1}{2} \frac{m_\pi^2}{(4\pi
  f_\pi)^2}\ln\frac{m_\pi^2}{\Lambda_3^2} \bigg\} (1+{c_s}\Delta)
  \quad\hbox{versus}\quad
  m_{ud}=c_1+c_2 m_\pi^2 +c_3 m_\pi^4 +c_4\Delta,
\ee
where $\Delta$ parameterises the deviation from the physical strange
quark mass. BMW have also applied cuts on the pion mass, by limiting
its maximum value to either~340 or 380\,\MeV.

\begin{figure}
\begin{center}
\includegraphics[width=12cm]{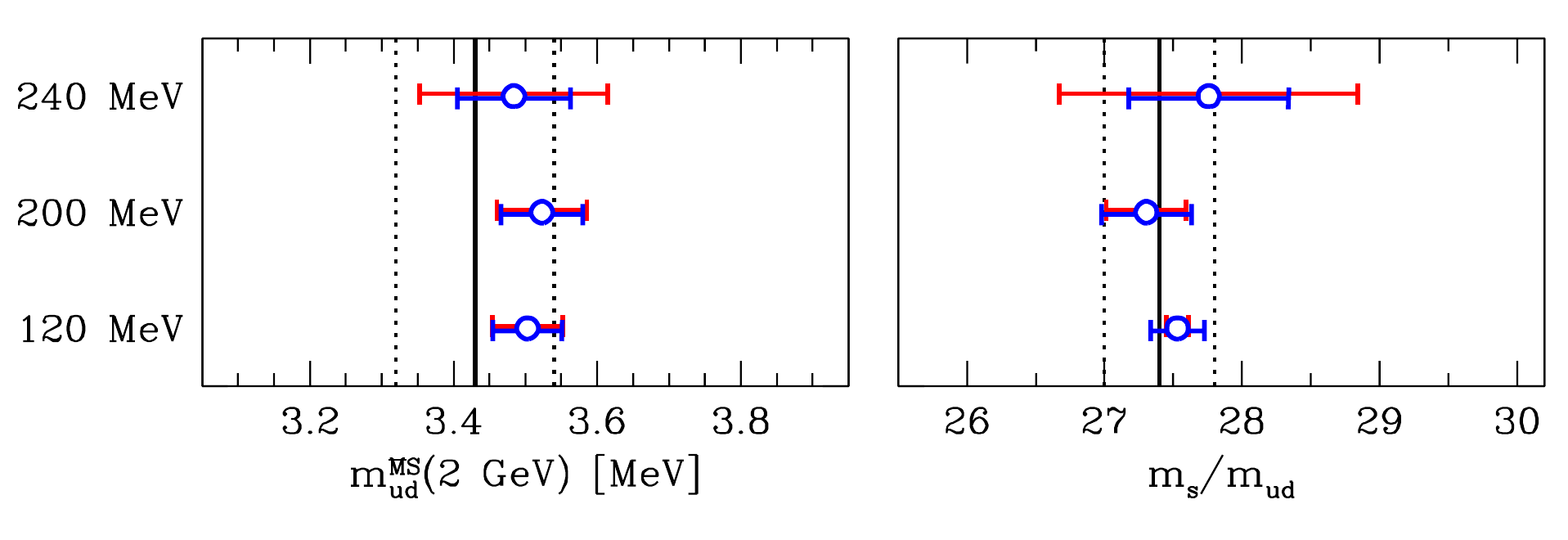}
\vspace{-0.5cm}
\caption{The effect of cuts applied to the lower limit of the pion
  mass interval on the light quark mass (left) and the ratio
  $m_s/m_{ud}$ (right)\,\cite{FodHoelb_priv}. Red error bars denote
  the systematic uncertainty. The vertical bands represent the global
  results of \protect\eq{eq:quark}.\label{fig:BMWcuts}}
\vspace{-0.5cm}
\end{center}
\end{figure}

Given the availability of ensembles with pion masses as low as
120\,\MeV, it is interesting to study the question of the impact of
interpolations in the pion mass, compared to relying on
extrapolations. In other words, how do the results for quark masses
and their uncertainties vary when the {\it lower} limit of the pion
mass interval to which the chiral fit is applied, increases gradually?
Figure\,\ref{fig:BMWcuts} shows the results for the light quark mass
$m_{ud}$ and the ratio $m_s/m_{ud}$ obtained after applying mass cuts
at $m_\pi=120,\,200$ and 240\,\MeV,
respectively\,\cite{FodHoelb_priv}. One clearly sees that the results
are quite stable and consistent within errors, indicating that chiral
extrapolations are under good control. Unsurprisingly, the error
increases for longer extrapolations. A look at
Table\,\ref{tab:BMWcuts} shows that the error budget is, in fact,
increasingly dominated by the fit {\it ansatz} when the minimum pion
mass is shifted to larger values. It would be helpful to discount the
possibility that these findings are obscured by lattice artefacts,
since below-physical pion masses have only been simulated for the
three coarsest lattice spacings. As a suggestion for an improved
future analysis of the BMW data, the effects of imposing lower mass
cuts should be investigated after the data have been extrapolated to
the continuum limit for fixed values of $m_\pi$.
\begin{table}
\begin{center}
\begin{tabular}{l|c c c| c c c c c c}
\hline\hline
cut & $m_{ud}$ & $\sigma_{\rm stat}$ & $\sigma_{\rm syst}$ & plateau & scale
    & fit form & mass cut & renorm. & cont. \\
\hline
120 MeV & 3.503 & 0.048 & 0.049 & 0.330 & 0.034 & 0.030 & 0.157 & 0.080 &
0.926 \\
200 MeV & 3.523 & 0.057 & 0.063 & 0.354 & 0.078 & 0.470 & 0.236 & 0.087
& 0.765 \\
240 MeV & 3.484 & 0.079 & 0.131 & 0.316 & 0.092 & 0.807 & 0.341 & 0.046 &
0.349 \\
\hline\hline
\end{tabular}
\caption{Error budget for the light quark mass $m_{ud}^{\msbar}(2\,\GeV)$
  after applying cuts to the minimum pion mass of 120,\,200 and 240\,\MeV. The
  first row corresponds to the original results in\,\cite{BMW2010-1}. The last
  six colums represent the relative contributions of individual systematic
  effects to the overall systematic error $\sigma_{\rm
    syst}$. \label{tab:BMWcuts}}

\end{center}
\vspace{-0.4cm}
\end{table}
%
%
The above discussion shows that simulations at or below the physical
pion mass allow for a systematic investigation into the quality of
chiral fits. In order to rightfully claim an overall accuracy of a few
percent in lattice estimates of the light quark masses, minimum pion
masses of $250\,\MeV$ appear to be sufficient.

\section{Systematics of chiral fits: meson decay constants} \label{sec:s3decay}

Masses and decays constants of pseudoscalar mesons belong to the set
of quantities whose dependence on the quark mass has been studied most
extensively. Chiral fits using lattice data and the expressions of
ChPT give access to the effective coupling constants (low-energy
constants -- LECs) of ChPT. These include the pion decay constant in
the chiral limit, $f$, the quark condensate $\Sigma$, and also some of
the LECs which enter at NLO in the chiral expansion (e.g. $\bar\ell_3$
and $\bar\ell_4$). Furthermore, the decay constants of the physical
pion and kaon, $f_\pi$, $\fK$, as well as the ratio $\fK/f_\pi$ are
determined in this way. While the individual decay constants are often
used to set the lattice scale, the ratio $\fK/f_\pi$ is important for
constraining the ratio $|V_{us}|/|V_{ud}|$ of CKM matrix
elements. Both aspects will be covered in this section. The discussion
here is restricted to lattice calculations in the $p$-regime.

\begin{figure}
\begin{center}
\includegraphics[height=7cm]{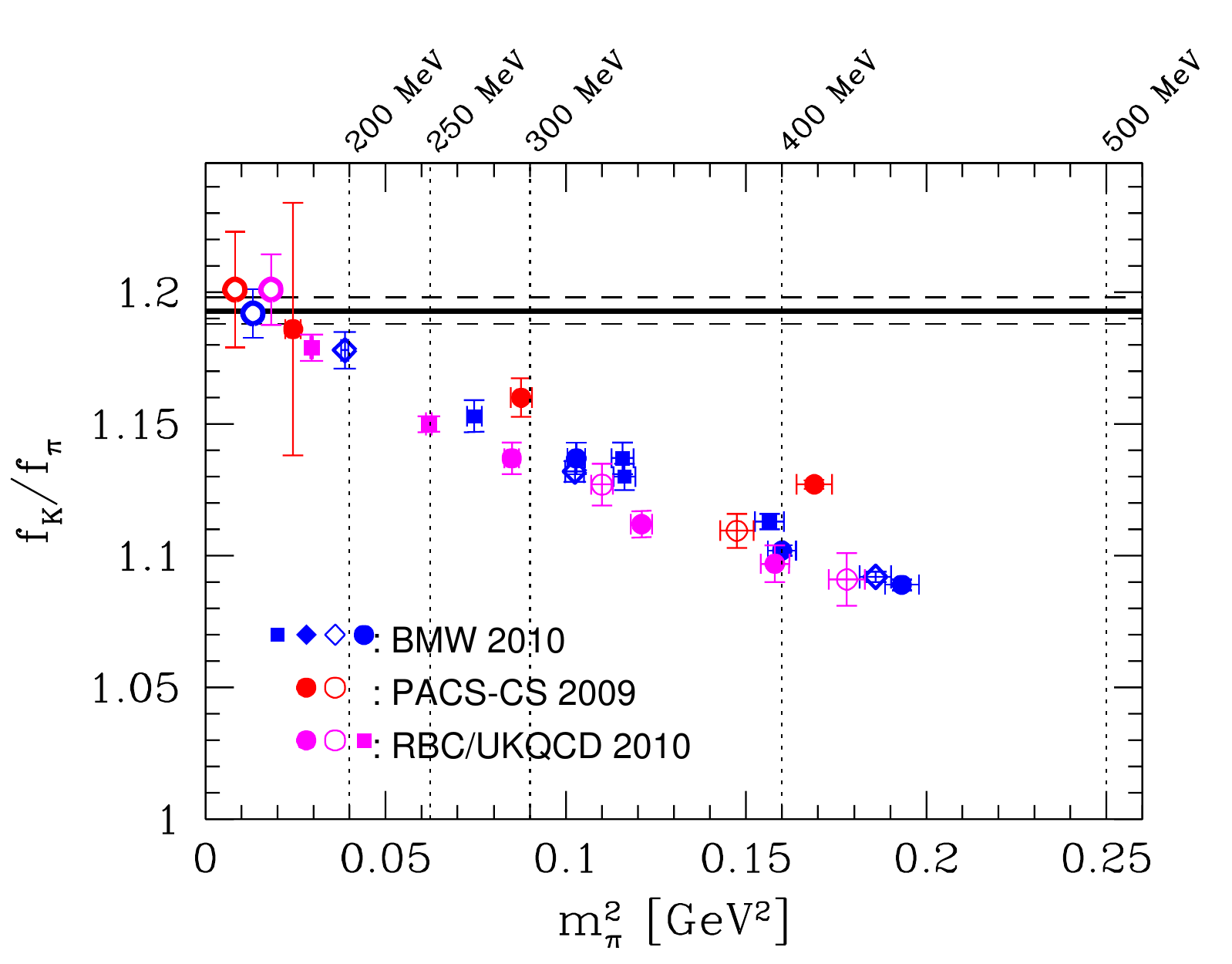}
\caption{\label{fig:fKfpi-fit} The ratio $\fK/f_\pi$ plotted versus
  the squared pion mass from
  refs.\,\cite{Durr:2010hr,Aoki:2008sm,RBC-UKQCD10A,Kellypriv}. Open
  circles denote the extrapolated values at the physical pion mass
  which are shifted for clarity. The horizontal lines represent the
  global estimate from the FLAG report\,\cite{FLAG2010} for
  $\Nf=2+1$.}
\vspace{-0.5cm}
\end{center}
\end{figure}

Lattice results for the ratio $\fK/f_\pi$ are in general quite stable
and consistent among different collaborations. Examples of chiral
extrapolations are shown in Fig.\,\ref{fig:fKfpi-fit}. The FLAG report
provides separate global estimates for QCD with $\Nf=2$ and $2+1$
flavours, i.e.
\be
   \fK/f_\pi = 1.193\pm0.005\quad(\Nf=2+1),\qquad
   \fK/f_\pi = 1.210\pm0.006\pm0.017\quad(\Nf=2).
\label{eq:fKfpi}
\ee
The result for $\Nf=2+1$ is based on
refs.\,\cite{Durr:2010hr,Bazavov:2010hj,Follana:2007uv}, while the
two-flavour result is identical to the value quoted
in\,\cite{Blossier:2009bx}.

As was pointed out by Marciano\,\cite{Marciano:2004uf}, a precise
lattice estimate of $\fK/f_\pi$ in conjunction with accurate
experimental measurements of the leptonic decay rates of $K_{\ell2}$
decays provides a stringent constraint on the ratio
$|V_{us}|/|V_{ud}|$ of CKM matrix elements, since
\be
   \frac{\Gamma(K\to\mu\bar\nu_\mu(\gamma))}
        {\Gamma(\pi{\to}e\bar\nu_e(\gamma))} \propto
   \frac{|V_{us}|^2}{|V_{ud}|^2}\,\frac{\fK^2\mK}{f_\pi^2 m_\pi}.
\ee
An additional constraint on $|V_{us}|$ is provided by the form factor
$f_{+}(q^2)$ which appears in the expression for the rate of the decay
$K\to\pi\ell\nu$. Lattice calculations for $f_{+}(q^2)$ are consistent
and equally precise compared with the effective field theory result of
ref.\,\cite{LeutwylerRoos} in which $f_{+}(0)$ was determined by
invoking the Ademollo-Gatto theorem\,\cite{Ademollo}, which states
that the corrections due to isospin and SU(3) flavour breaking are
second order. Global lattice estimates for $f_{+}(0)$ are quoted
in\,\cite{FLAG2010} as
\be
    f_{+}(0)=0.9597\pm0.0038\quad(\Nf=2+1),\qquad
    f_{+}(0)=0.9604\pm0.0075\quad(\Nf=2).
\label{eq:fplus}
\ee
The current accuracy of lattice results for $\fK/f_\pi$ and $f_{+}(0)$
allows for a precision test of first-row unitarity of the CKM matrix,
i.e.
\be
  |V_{ud}|^2+|V_{us}|^2+|V_{ub}|^2=1.
\ee
The contribution from the $b$-quark can be dropped, since
$|V_{ub}|^2=\rmO(10^{-5})$, which is below the relevant level of
accuracy in the following discussion. The experimental branching
fractions are
\be
    |V_{us}| f_{+}(0) = 0.2163(5), \qquad
    \left|\frac{V_{us}{\fK}}{V_{ud}{f_\pi}}\right| = 0.2758(5).
\ee
Combining these values with the global lattice estimates for $\Nf=2+1$
in eqs.\,(\ref{eq:fKfpi}) and~(\ref{eq:fplus}) yields
\be
   |V_{ud}|^2+|V_{us}|^2 = 1.002\pm0.015.
\ee
The precision of this test can be considerably enhanced by including
another experimental constraint, namely the determination of $V_{ud}$
from super-allowed nuclear $\beta$-decays. Using the lattice result
for the ratio $\fK/f_\pi$ which fixes $|V_{us}|/|V_{ud}|$
gives\,\cite{FLAG2010}
\be
   |V_{ud}|^2+|V_{us}|^2 = 0.9999\pm0.0006.
\ee
In this way, first-row unitarity is confirmed with per-mille accuracy,
using experimental information and lattice estimates alone. The
unitarity test is equally precise if the lattice estimate is provided
by $f_{+}(0)$ instead of $\fK/f_\pi$.

The above discussion suggests that lattice calculations of
pseudoscalar meson decay constants are under very good control.
Table\,\ref{tab:r0decay} contains a compilation of recent estimates
for decay constants, the hadronic radii $r_0$ and $r_1$ and certain
combinations thereof. Although the ratio $\fK/f_\pi$ is consistent
among different calculations within the quoted errors, this is not
necessarily true for the absolute values of decay constants and
hadronic radii. On the assumption that there are no significant
differences between two- and three-flavour QCD within the presently
quoted errors, one finds that the $r_0$ determination from
ETMC\,\cite{Blossier:2009bx} (which uses the physical value of $f_\pi$
to set the scale) contradicts the estimate quoted by
RBC/UKQCD\,\cite{RBC-UKQCD10A}. On the other hand, RBC/UKQCD, who
determine the lattice scale from the mass of the $\Omega$ baryon, find
a value for $f_\pi$ which is smaller than the experimental
value. Similar observations apply to the results for $\fK$, $r_1$ and
${\fK}r_1$ quoted by RBC/UKQCD and MILC\,\cite{Bazavov:2009bb}.

\begin{table}
\begin{center}
{\small
\begin{tabular}{l c c c c l l l l}
\hline\hline
 & scale & $f_{\pi}$ &  $\fK$ & $\fK/f_{\pi}$ & $r_{0}\,[\fm]$ &
           $r_{1}\,[\fm]$ & $\fK r_{0}$ & $\fK r_{1}$  \\ 
\hline
RBC /  & & & & & & & & \\
UKQCD\rb{\cite{RBC-UKQCD10A}} & \rb{$m_\Omega$} & \rb{124(5)} &
\rb{149(5)} & \rb{1.204(26)} & \rb{0.487(9)} & \rb{0.333(9)} & \rb{0.368(9)} &
\rb{0.251(10)} \\[0.1cm]
PACS-CS\cite{Aoki:2008sm} & $m_\Omega$ & 134(4) & 159(3) & 1.189(20) &
   0.492($^{10}_{~6}$) & & 0.397($^{11}_{~9}$) & \\[0.1cm]
MILC\,\cite{Bazavov:2009bb} & $f_\pi$ & ./. & 157($^{1}_{3}$) &
1.197($^{~7}_{13}$) & & 0.311($^{3}_{8}$) & & 0.246(5) \\[0.1cm]
ETMC\,\cite{Blossier:2009bx} & $f_\pi$ & ./. & 158(2) & 1.210(18) &
0.438(10) & & 0.351(10) \\
\hline\hline
\end{tabular}
}
\end{center}
\caption{Results for decay constants and the hadronic radii $r_0$ and $r_1$
  from simulations with $\Nf=2+1$ dynamical flavours (RBC/UKQCD, PACS-CS,
  MILC) and $\Nf=2$ (ETMC). The renormalisation of the axial current
  in\,\cite{Aoki:2008sm,Bazavov:2009bb} was done
  perturbatively.\label{tab:r0decay}} 
\end{table}

One may suspect that the observed differences in the absolute values
of $\fK$ and $f_\pi$ are linked to the normalisation of the axial
current. In fact, only the correctly normalised matrix elements can be
expected to approach the continuum limit with a rate proportional to
the leading lattice artefacts. For quantities such as $r_0$ and $r_1$
the situation is not much better, because little is known about the
chiral behaviour one is to expect. In view of the importance of decay
constants and hadronic radii for the overall scale setting, one must
make an effort to understand the observed differences. Several
collaborations have reported new results for these quantities at this
conference
\cite{CKelly_lat11,Lightman_lat11,Scholz_lat11,VdW_lat11,Marina_lat11,Bjoern_lat11}.

\begin{figure}
\begin{center}
\leavevmode
\includegraphics[height=5cm]{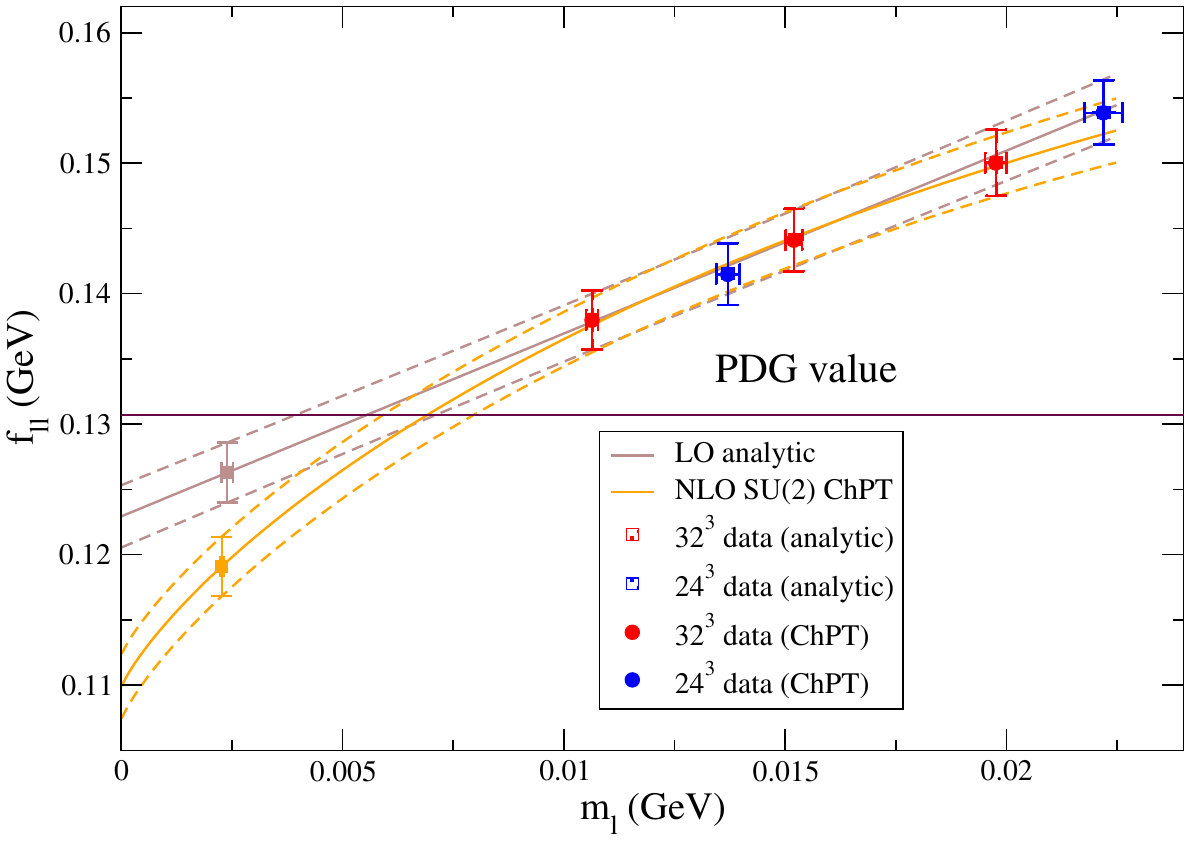}
\hfill
\includegraphics[height=5.52cm]{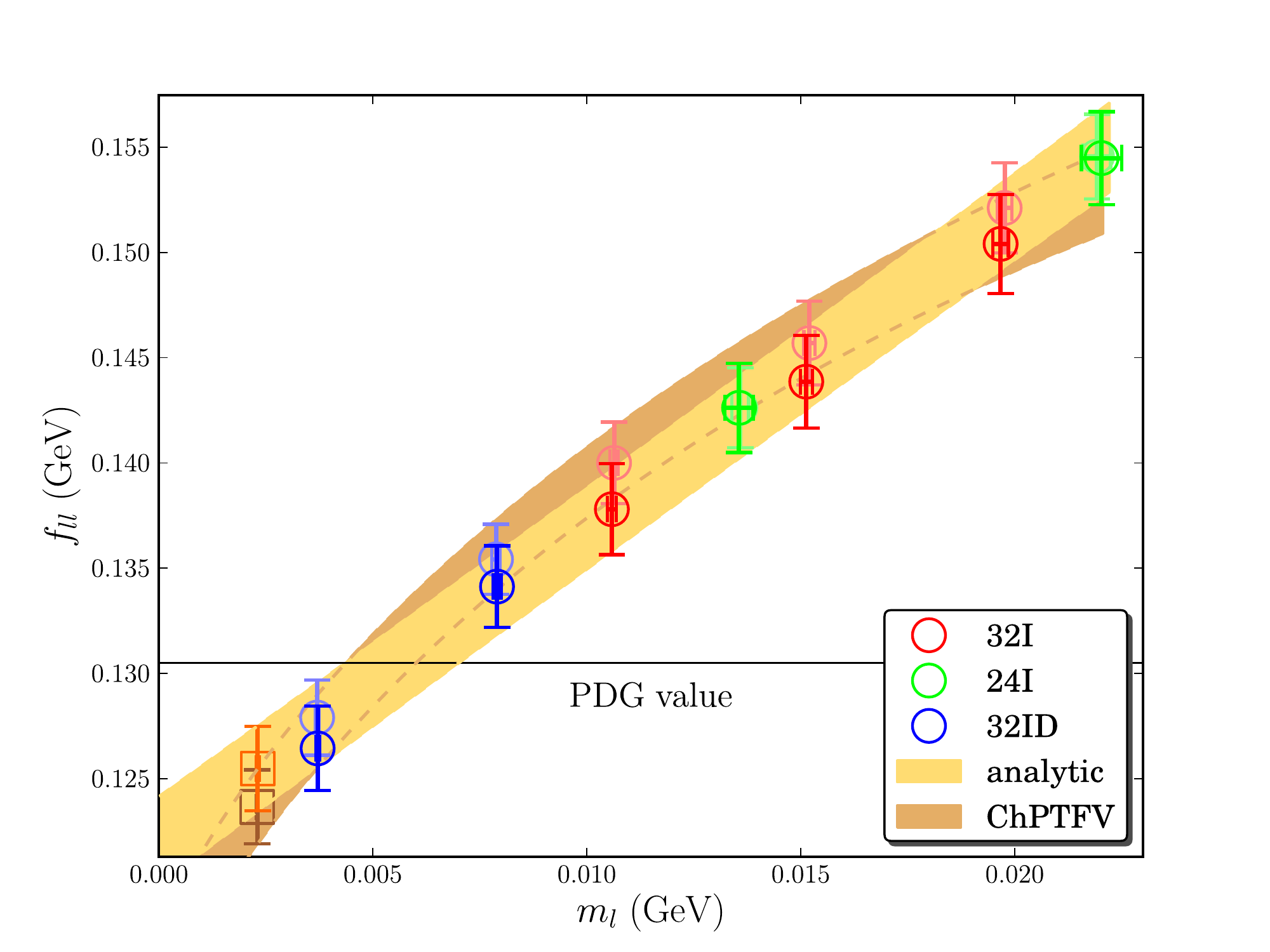}
\caption{\label{fig:fpiRBC} Chiral extrapolations of $f_\pi$ from
  RBC/UKQCD. The left panel shows the comparison of SU(2) ChPT with an
  analytic {\it ansatz}, using a minimum pion mass of 290\,\MeV at
  $a\approx0.11\,\fm$\,\cite{RBC-UKQCD10A}. In the right panel two
  additional points with $m_\pi=170$ and $250\,\MeV$ at
  $a\approx0.014\,\fm$ have been included\,\cite{CKelly_lat11}.}
\vspace{-0.5cm}
\end{center}
\end{figure}

RBC/UKQCD have investigated why the individual decay constants $f_\pi$
and $\fK$ are lower than experiment, while their ratio agrees with
other simulations. To this end they have supplemented their existing
data sets by two more ensembles with pion masses of~250 and
170\,\MeV\,\cite{CKelly_lat11}. In order to keep ${m_\pi^{\sf
min}}L>4$ for $L/a=32$, lower pion masses could be simulated at the
expense of having to use coarser lattice spacings
($a\approx0.14\,\fm$) compared
to\,\cite{RBC-UKQCD10A}. Figure\,\ref{fig:fpiRBC} shows the impact of
the additional data points on the chiral extrapolation of
$f_\pi$. Clearly, the ambiguity associated with the {\it ansatz} for
the chiral behaviour is reduced: Extrapolations based either on SU(2)
ChPT or on a Taylor expansion produce results for $f_\pi$ which agree
very well within errors. However, while the ratio $\fK/f_\pi$ is
consistent with the earlier result, the new and preliminary value of
$f_\pi=125(2)(3)\,\MeV$ still appears to be smaller than the
experimental value. It should be kept in mind that the entire range of
pion masses which RBC/UKQCD have access to, involves different
lattice spacings. The systematics of chiral fits can only be
investigated reliably provided that the dependence on the lattice
spacing is well understood.

\begin{figure}
\unitlength 1cm
\begin{picture}(16,7)(0,0)
\put(0.5,2.5){\includegraphics[height=4.0cm]{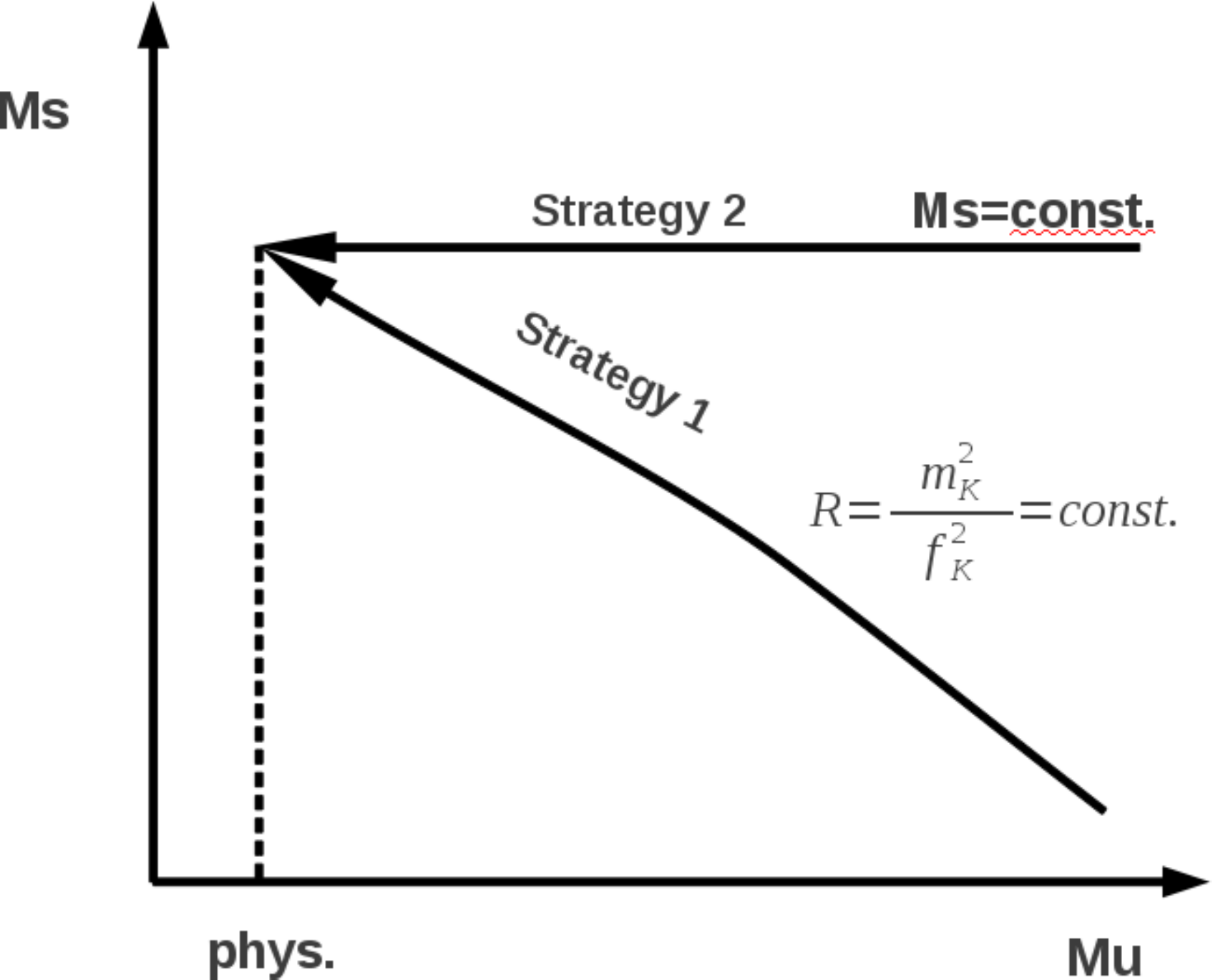}}
\put(6.5,7.0){\includegraphics[angle=270,width=8.0cm]{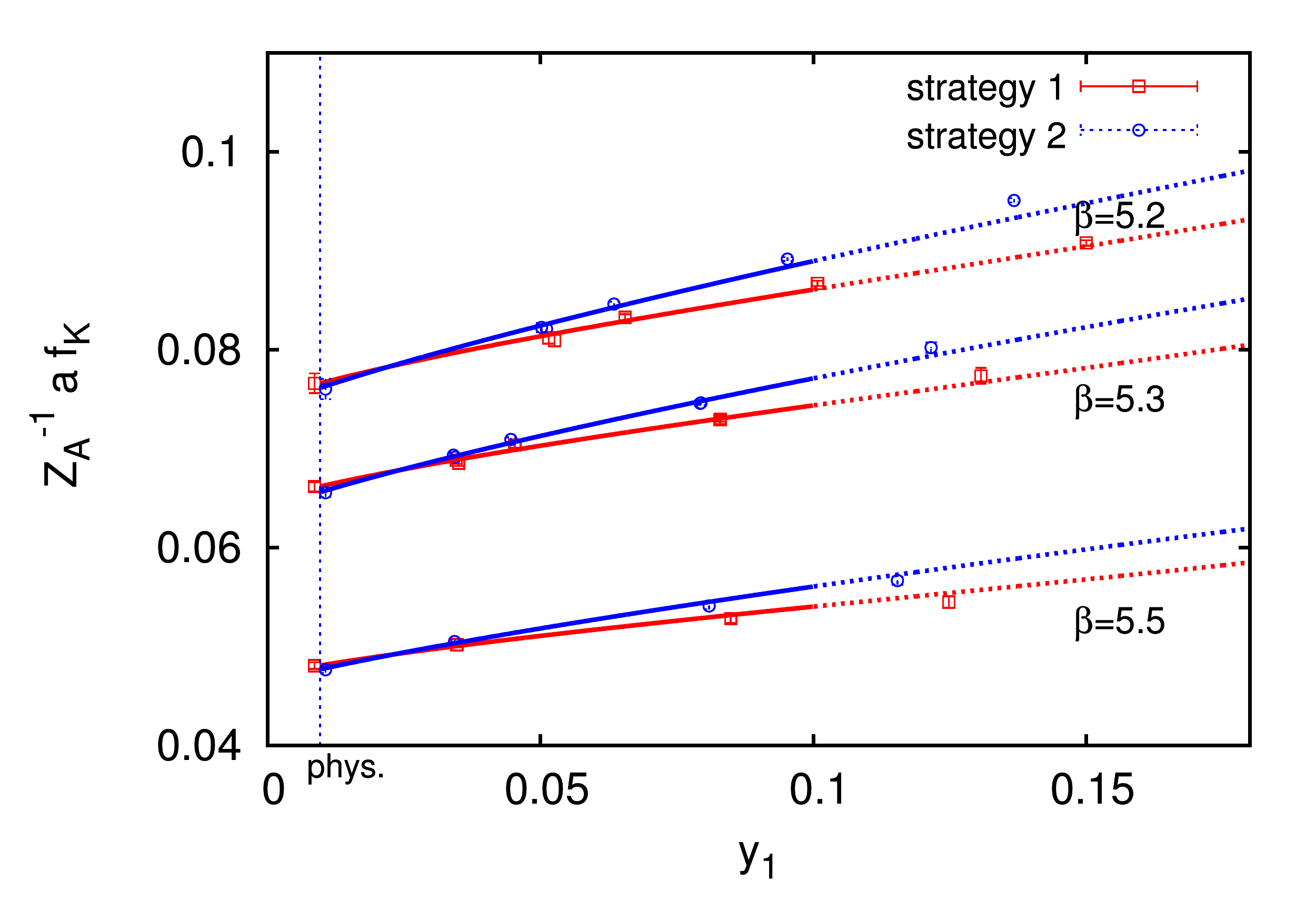}}
\end{picture}
\vspace{-2.2cm}
\caption{Chiral extrapolation of $\fK$ by ALPHA\,\cite{Marina_lat11}. Left:
  schematic view of the paths of two different extrapolations in the
  $(m_{ud},m_2)$-plane. Right: chiral extrapolations of $a\fK/\za$ in the variable
  $y_1\equiv m_\pi^2/8\pi^2\fK^2$ for three different bare
  couplings.\label{fig:marina}}
\end{figure}

The consistency of different chiral fit {\it ans\"atze} has also been studied
by the ALPHA Collaboration for $\rmO(a)$ improved Wilson
quarks\,\cite{Marina_lat11}. In order to check the robustness of the
extrapolation of $\fK$ to the physical pion mass, they have compared two
different fit strategies. Denoting the hopping parameters by $\kappa_1$ and
$\kappa_2$, where $\kappa_1=\kappa_{\rm sea}$, the first strategy amounts to
adjusting $\kappa_2$ until $\mK^2/\fK^2$ is equal to the experimental
value. Repeating this for each sea quark mass defines, at leading order, a
sequence of data points with $m_s+m_{ud} =\rm const.$, which can be
extrapolated to the physical pion mass using the expressions from partially
quenched ChPT. In the second strategy the strange quark mass is held fixed: at
each value of $\kappa_{\rm sea}$, the hopping parameter $\kappa_2$ is tuned
such that the PCAC mass reproduces the fixed value of $\mu\equiv m_s$. The
resulting values of $\fK$ can then be extrapolated in the pion mass using
SU(2) ChPT at NLO. A schematic view how the physical point in the
$(m_{ud},m_s)$-plane is approached with the two strategies is shown in
Fig.\,\ref{fig:marina} (left).  The right panel of the figure compares the
chiral extrapolation of the (bare) $\fK$ for three different values of
$\beta$. Even though they are based on quite different variants of ChPT, both
strategies converge to the same values at the physical pion mass, which
enhances the credibility of the chiral fits. After taking the renormalisation
of the axial current into account the results can be used to set the lattice
scale using $\fK$. In this way ALPHA find that the three $\beta$-values
correspond to $a\approx0.075, 0.066$ and $0.049\,\fm$, respectively. These
preliminary results agree well within errors with the scale determination via
the mass of the $\Omega$ baryon\,\cite{Georg_lat11} on the same ensembles.

\begin{figure}
\begin{center}
\leavevmode
\includegraphics[height=5.0cm]{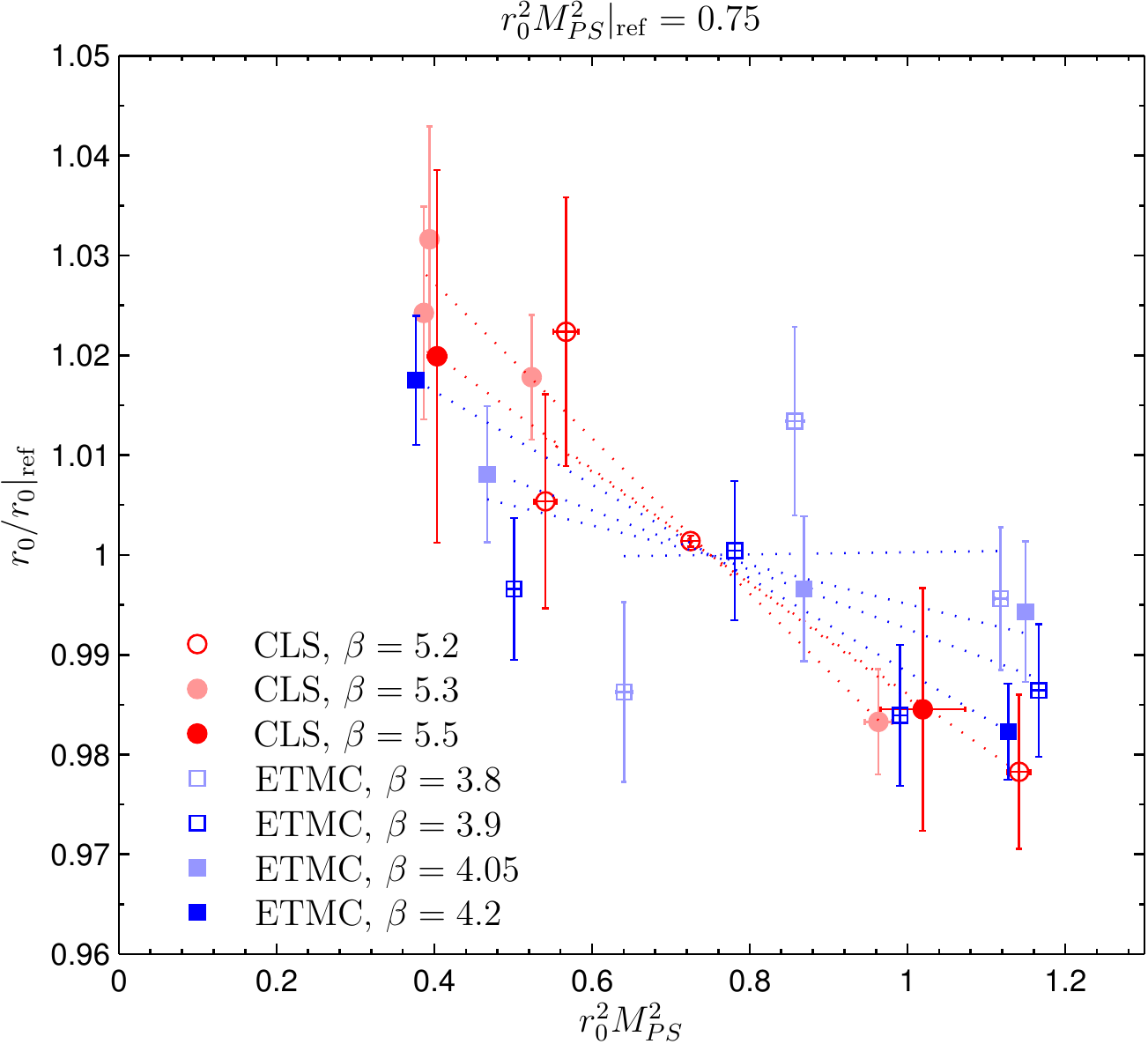}
\hspace{2cm}
\includegraphics[height=4.83cm]{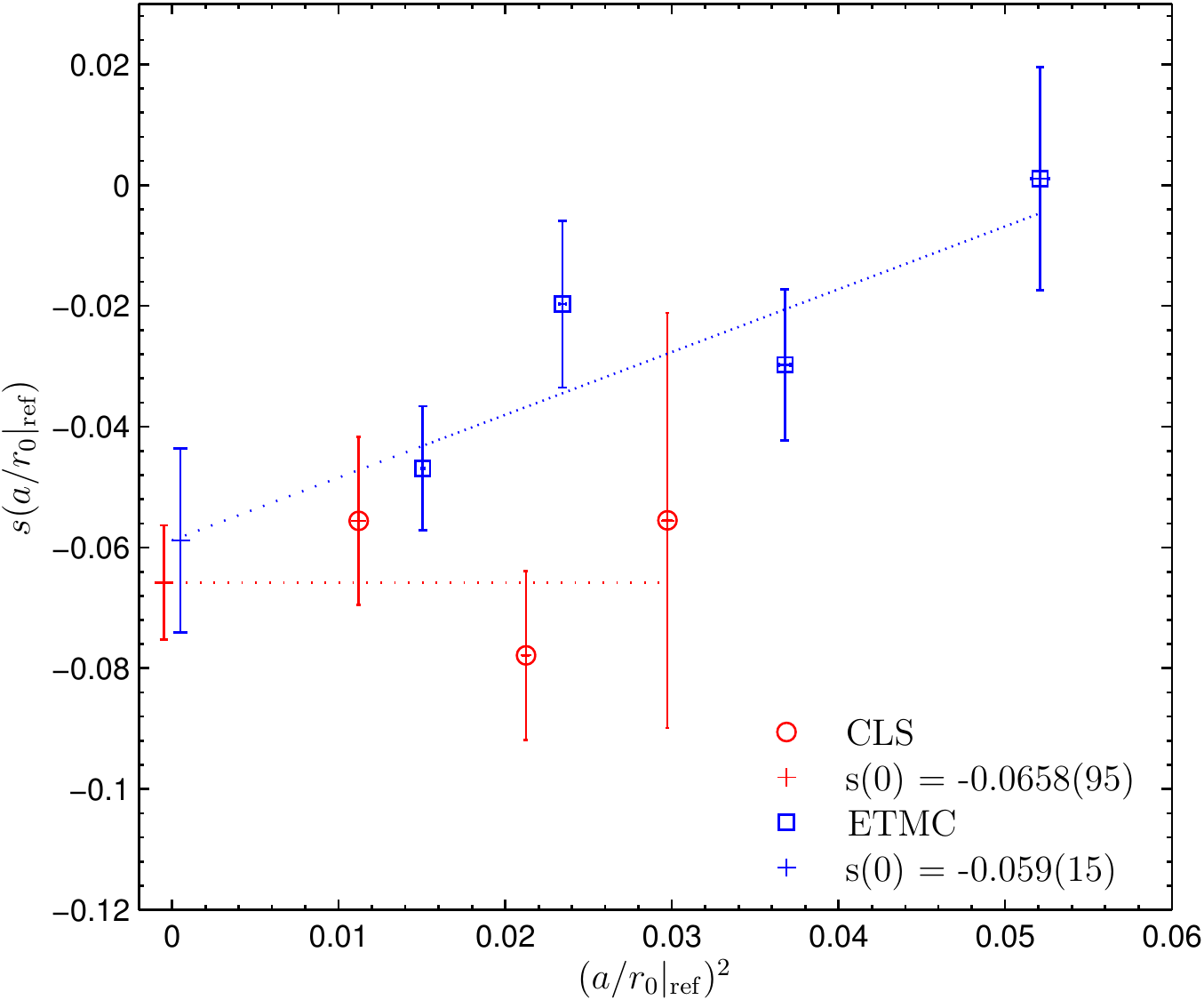}
\caption{Left: pion mass dependence of $r_0/\left.r_0\right|_{\rm ref}$
  computed on the CLS ensembles\,\cite{Bjoern_lat11} (red symbols) compared to
  the results from ETMC\,\cite{ETMC_r0} (blue symbols). The quantity
  $r_0|_{\rm ref}$ denotes the value at a reference pion mass. Right:
  continuum extrapolation of the slope parameter~$s$ in
  eq.\,(\protect\ref{eq:r0fit}).\label{fig:r0}}
\vspace{-0.5cm}
\end{center}
\end{figure}

Decay constants are often used to calibrate the hadronic radii $r_0$ and
$r_1$, by determining combinations such as ${f_\pi}r_0$ at the physical pion
mass in the continuum limit. Chiral extrapolations of $r_0$ and $r_1$ are
usually based on the assumption that they can be described by a polynomial in
$m_\pi^2$. A detailed investigation of the chiral behaviour of $r_0$ was
presented at this conference\,\cite{Bjoern_lat11} (see also
ref.\,\cite{Rainer_lat03}). In Fig.\,\ref{fig:r0} the pion mass dependence of
$r_0$ computed on the CLS ensembles with $\rmO(a)$ improved Wilson quarks is
compared to the results by ETMC\,\cite{ETMC_r0}. As can be seen from the left
panel of the figure, $r_0$ indeed shows a linear dependence on the squared
pion mass at a given value of the bare coupling. By performing a linear fit
according to
\be
 r_0/\left.r_0\right|_{\rm ref} = A + s\times ({m_\pi}r_0)^2,
\label{eq:r0fit}
\ee
one can determine the slope parameter~$s$, which is found to depend
quite strongly on the lattice spacing in the case of ETMC. The right
panel of Fig\,\ref{fig:r0} shows the extrapolation of~$s$ to
$a=0$. The rate with which this quantity approaches the continuum
limit is indicative of the size of lattice artefacts of
$\rmO({m_q}a^2)$. While the slope~$s$ from the two collaborations
agrees very well in the continuum limit, there are sizeable
corrections proportional to ${m_q}a^2$ for twisted-mass QCD. This
discussion serves as a reminder that the problem of mass-dependent
lattice artefacts, which are formally of $\rmO(a^2)$ must be
addressed, since they will affect the determination of ${f_\pi}r_0$
and hence the calibration of $r_0$.

\section{Nucleon axial charge} \label{sec:s4axial}

While hadronic uncertainties in the meson sector could be brought well
under control, the situation for baryonic quantities is much less
satisfactory.  Despite many years of dedicated effort, lattice results
for nucleon form factors or moments of structure functions fail to
reproduce the experimental values within the quoted
uncertainties\,\cite{renner_lat09,dina_lat10}. A prominent example is
the axial charge, $\gA$, of the nucleon. Lattice simulations using
pion masses $m_\pi\;\gtaeq\;250\,\MeV$ typically underestimate this
quantity by $10-15$\,\%. Even more worrisome is the observation that
the gap is stable, i.e. the data show little if no tendency to
approach the physical value as the pion mass is decreased. There is a
broad concensus that uncontrolled systematic effects must be held
responsible.

The axial charge is an ideal observable to study lattice systematics
for baryonic quantities: It is defined in terms of a transition matrix
element at zero momentum transfer and hence the underlying kinematics
is very simple. Second, it can be determined without the evaluation of
quark-disconnected diagrams.  Among the common sources of systematic
error are lattice artefacts, the related issue of the correct
normalisation of the axial current, and the influence of finite-volume
effects, which are known to be larger for baryonic systems. An obvious
question is whether the chiral behaviour is sufficiently controlled in
the calculations performed so far, or whether much smaller pion masses
are required in order to make contact with the experimental value.
Another issue which has received quite some attention recently, is the
possible contamination of baryonic three-point correlation functions
by contributions from higher excited states. This seems plausible,
since the noise-to-signal ratio in baryonic correlation functions is
much worse than for mesons. Thus, one cannot firmly rule out the
possibility that excited state contributions are still present within
the relatively short Euclidean time interval before the signal is
lost.

The theoretical foundations of baryonic ChPT, which is used to
constrain the chiral behaviour of $\gA$, are unfortunately on a weaker
footing compared to the mesonic sector. Since the mass gap between the
nucleon and the nearest resonance, i.e. the $\Delta$, is much smaller
than the mass scale defined by the nucleon itself, it is difficult to
define a consistent chiral counting scheme. The established formalisms
include Heavy Baryon ChPT\,\cite{HBChPT}, the infrared regularisation
of loop integrals\,\cite{BecherLeut} and the related extended on-mass
shell regularisation\,\cite{EOMS_MZ}, some of which have been carried
to high orders in the expansion. Another approach is the so-called
small-scale expansion (SSE)\,\cite{SSEChPT}, in which the
nucleon-$\Delta$ splitting is treated as a small parameter and
included in the chiral power counting in the framework of Heavy Baryon
ChPT. One severe drawback for the interpretation of lattice data is
the large number of coupling terms, each of which carries a low-energy
constant. Some of these LECs can be constrained from phenomenology,
but unless one has access to extremely detailed information from
lattice simulations deeply in the chiral regime, it seems impossible
to determine the full set.

The bare value of $\gA$ can be extracted from a suitable ratio of two-
and three-point functions. In the simplest case, i.e. when the same
operators are used to create and annihilate the nucleon, the
expression reads
\be
  R_{\rm{A}}(t,t_s) = \frac{C_3^{\rm{A}}(t,t_s)}{C_2(t_s)}\;
  \stackrel{t,(t_s-t)\gg0}{\longrightarrow}\; \gA^{\rm{bare}}
  +\rmO(\rme^{-\Delta_{\rm{N}}t})   +\rmO(\rme^{-\Delta_{\rm{N}}(t_s-t)}).
\label{eq:gA}
\ee
Here $t_s$ denotes the Euclidean time separation between the initial
and final nucleons, while the axial current is inserted at time $t$
with $0\leq t\leq t_s$. Due to the rapidly increasing statistical
noise, typical values of $t_s$ are of order 1\,\fm, and thus the
correlation functions must reach their asymptotic behaviour for
separations $t,\,(t_s-t)\,\lesssim\,0.5\,\fm$. Since the gap
$\Delta_{\rm{N}}$ between the nucleon and its first excitation is
expected to scale like $\Delta_{\rm{N}}\sim 2m_\pi$ in the chiral
regime, it is clear from\,\eq{eq:gA} that corrections from excited
states become increasingly important as the physical pion mass is
approached.

Several collaborations have investigated the issue of excited state
contaminations recently. Using the CLS configurations generated with $\Nf=2$
flavours of $\rmO(a)$-improved Wilson fermions, the Mainz group has calculated
baryonic three-point functions for several different source-sink separations
$t_s$\,\cite{Mainz_FFs}. After computing the so-called ``summed
insertions''\,\cite{Maiani1987} according to
\be
   S_{\rm{A}}(t_s) = \sum_{t=0}^{t_s} R_{\rm{A}}(t,t_s)\;
  \stackrel{t_s\gg0}{\longrightarrow}\; \hbox{const.}+\gA^{\rm{bare}}t_s
  +\rmO(t_s\rme^{-\Delta_{\rm{N}}t_s}),
\label{eq:summation}
\ee
they determine the axial charge from the linear slope of $S_{\rm{A}}$
in the source-sink separation $t_s$. Since $t_s > t,\,(t_s-t)$ by
construction, it is clear that the corrections due to excited state
contamination in \eq{eq:summation} are parametrically more strongly
suppressed than for the simple ratio $R_{\rm{A}}(t,t_s)$. Other ways
to address excited state contamination include the use of
multi-exponential fits\,\cite{green_lat11} and systematic studies of
the dependence of $\gA$ on the source-sink separation for $t_s$ as
large as $1.9\,\fm$\,\cite{ETMC_exc2011}.
\begin{figure}
\begin{center}
\leavevmode
\includegraphics[height=6.2cm]{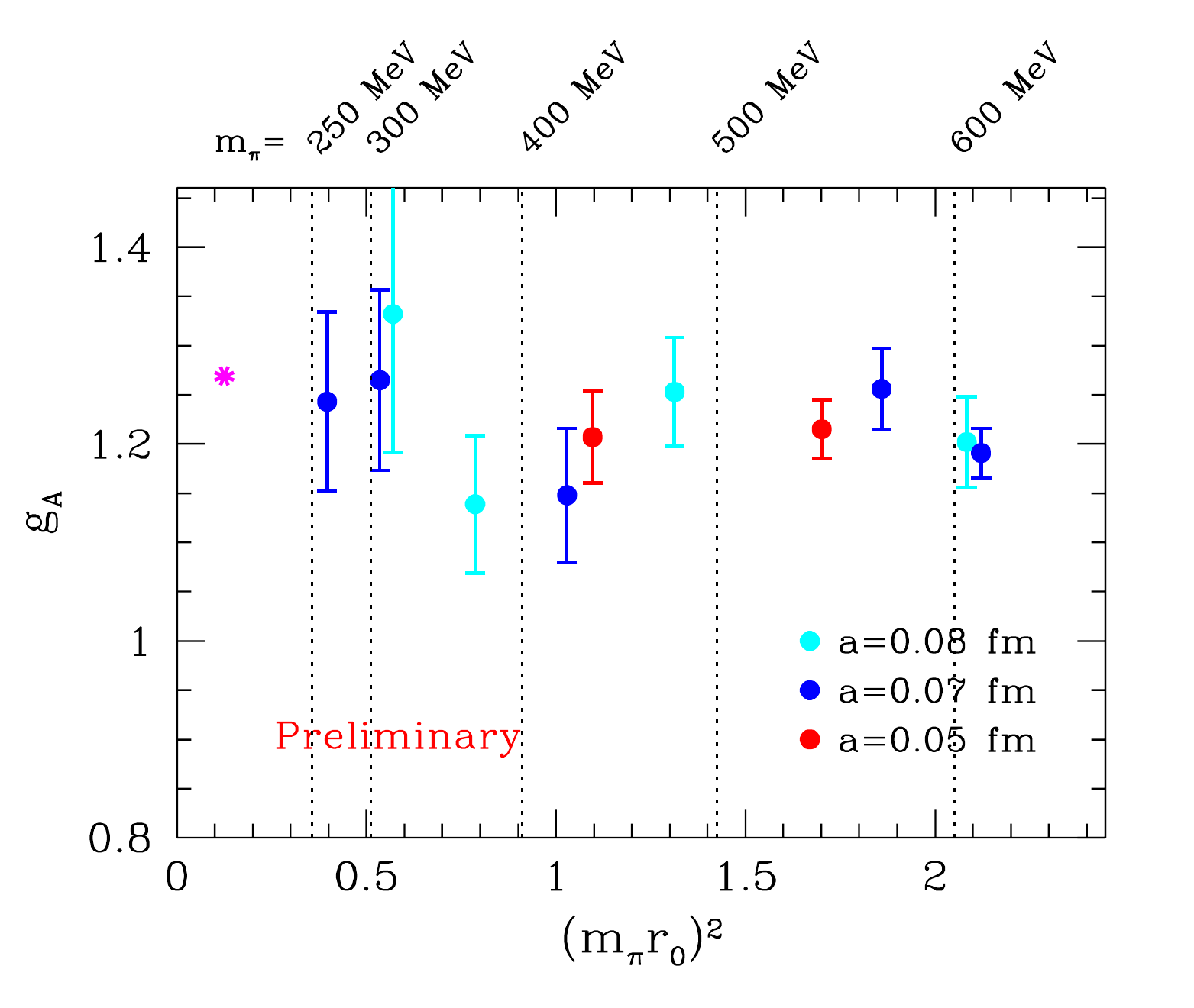}
\includegraphics[height=5.6cm]{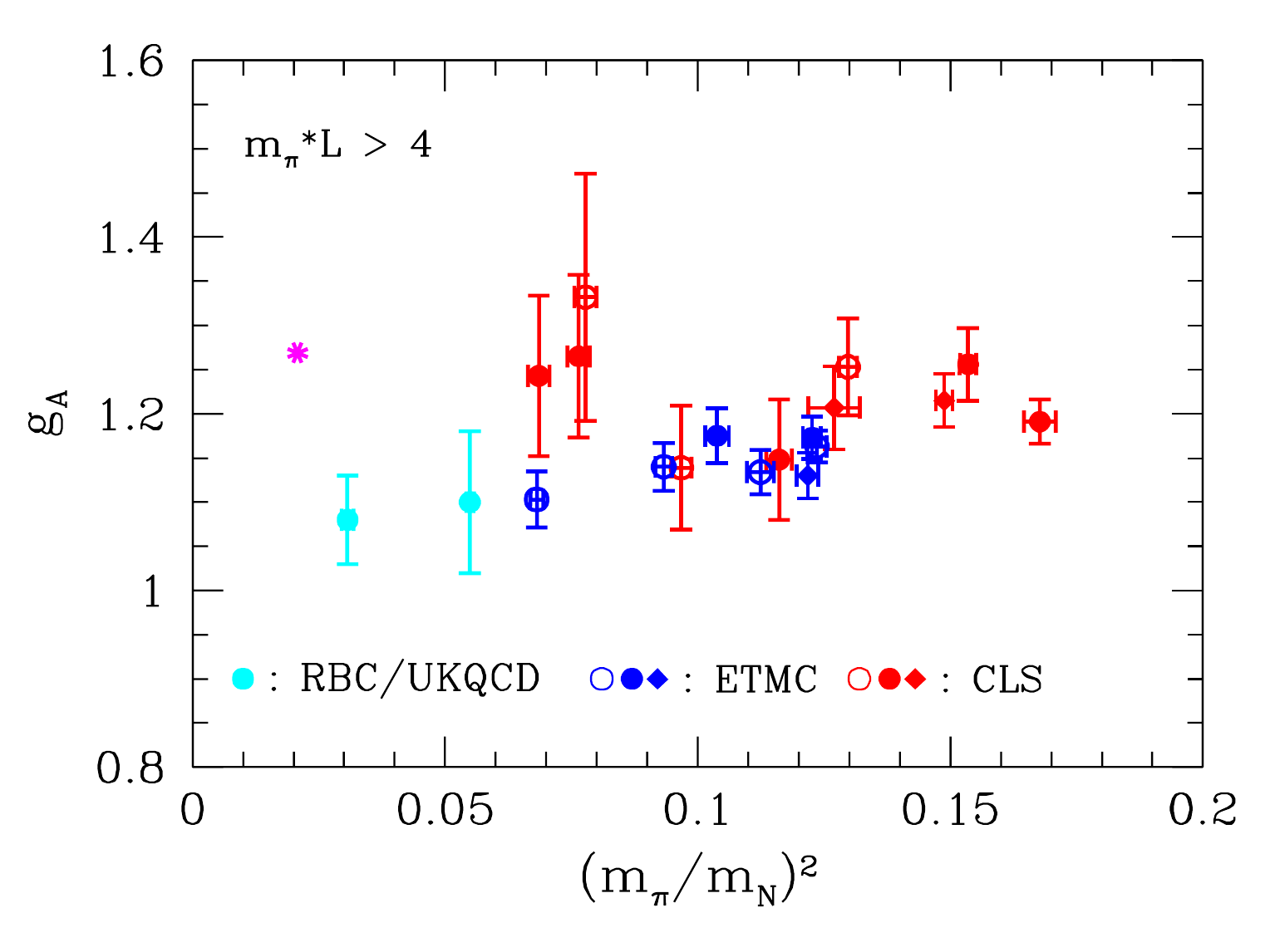}
\caption{\label{fig:gA} Left: preliminary results for $\gA$ computed
  on the CLS configurations using the technique of summed
  insertions\,\cite{Mainz_FFs}. Right: compilation of recent results
  for $\gA$ from refs.\,\cite{ETMC_gA2010,Mainz_FFs,ohta_lat11}. The
  experimental result for $\gA$ is represented by the pink asterisk.}
\vspace{-0.5cm}
\end{center}
\end{figure}
All these efforts do not allow for a firm conclusion at this
stage. The preliminary results by the Mainz group (left panel of
Fig.\,\ref{fig:gA}) suggest that summed insertions lead to a better
agreement with the experimental result for
$\gA$. ETMC\,\cite{ETMC_exc2011} report the absence of a bias in $\gA$
at $m_\pi=380\,\MeV$ but see some evidence for a distortion in the
case of $\langle{x}\rangle_{u-d}$. For the latter quantity,
LHPC\,\cite{green_lat11} can confirm that multi-exponential fits lead
to a better agreement with experiment as the pion mass is lowered,
albeit with a larger statistical error.

New results on $\gA$ by the RBC/UKQCD Collaborations were presented at
this conference\,\cite{ohta_lat11}, computed on the set of gauge
configurations which included the recently added lighter pion masses
discussed in section\,\ref{sec:s3decay}. They report a stable gap
between their preliminary results and the experimental value of $\gA$
across the entire mass range. The favoured explanation offered by
RBC/UKQCD is that the discrepancy is a result of finite-volume effects
rather than excited state contamination. A compilation of recent
results for $\gA$\,\cite{ETMC_gA2010,Mainz_FFs,ohta_lat11} is shown in
the right panel of Fig.\,\ref{fig:gA}, where the chiral behaviour is
compared among different groups after applying the cut
$m_{\pi}L>4$. Despite the very different systematics concerning the
discretisation of the quark action, the values of the lattice spacing
and the numerical procedures to extract $\gA$ from the measured
correlation functions, the results are broadly consistent with each
other. However, it would appear that those calculations which address
the issue of excited state contamination compare more favourably with
the experimental value.

\section{Summary and conclusions} \label{sec:s5summ}

Chiral extrapolations have been a persistent source of systematic
errors in lattice calculations. In this review I have tried to assess
the reliability of chiral extrapolations in order to investigate
whether the claimed accuracy of lattice results for several
phenomenologically interesting quantities is justified. The emergence
of simulation data around the physical pion mass was crucial, since it
allowed for a systematic study into the effects of replacing the
chiral extrapolation by an interpolation.

Pion masses in the range of $250 - 400\,\MeV$ appear to be sufficient
to guarantee that lattice estimates for the light quark masses can be
obtained with overall uncertainties at the level of a few
percent. Similarly, the chiral behaviour of the ratio $\fK/f_\pi$ is
under good control. The latter allows for a precise determination of
the ratio $|V_{us}/V_{ud}|$ and for a test of first-row unitarity with
permille accuracy, based on lattice results and experiment alone. For
individual decay constants, however, small inconsistencies among
different calculations remain and must be resolved. The separation of
lattice artefacts from systematic effects associated with the
description of the chiral behaviour must be improved not only for
$\fK$ and $f_\pi$ but also for quantities such as $r_0$.

In spite of these successes, one finds that lattice calculations for
the axial charge are still in an unsatisfactory state, since the
chiral behaviour of $\gA$ is clearly obscured by systematic
effects. With the presently available data it is difficult to decide
whether one single cause is chiefly responsible or whether it is a
convolution of finite-volume effects, excited state contamination and
lattice artefacts. Due to the unfavourable signal-to-noise ratio
of baryonic correlation functions it is likely that this can only be
resolved via an enormous increase in statistics.

\medskip
\par\noindent{\bf Acknowledgments:} I wish to thank O. B\"ar, Z. Fodor,
Ch. H\"olbling, C. Kelly, J. Laiho, B. Leder, M. Lightman, M. Marinkovic,
G. M\"unster, S. Ohta, R. Sommer, N. Tantalo, and G. Schierholz, for sending
new material prior to the conference and for valuable discussions. I am
grateful to Michele Della Morte, Georg von Hippel and Rainer Sommer for a
careful reading of the manuscript.

\end{document}